\documentclass[showpacs,twocolumn,prl,superscriptaddress,notitlepage,nofootinbib]{revtex4-2}
\usepackage[dvips]{graphicx}
\usepackage{amsmath}
\usepackage{amsmath,amssymb,amsthm,mathrsfs,amsfonts,dsfont}
\usepackage{subfigure, epsfig}
\usepackage{braket}
\usepackage{bm}
\usepackage{enumerate}
\usepackage{color}
\usepackage[normalem]{ulem}
\usepackage{graphicx}
\usepackage{dcolumn}
\usepackage{bm}
\newcommand{\css}{Co$_3$Sn$_2$S$_2$}
\usepackage{hyperref}
\begin{document}
\title{The magnetic structure of the topological semimetal Co$_3$Sn$_2$S$_2$}
\author{Jian-Rui Soh}
\affiliation{Institute of Physics, Ecole Polytechnique Fédérale de Lausanne (EPFL), CH-1015 Lausanne, Switzerland}
\author{ChangJiang Yi}
\affiliation{Beijing National Laboratory for Condensed Matter Physics, Institute of Physics, Chinese Academy of Sciences, Beijing 100190, China}
\author{Ivica Zivkovic}
\affiliation{Institute of Physics, Ecole Polytechnique Fédérale de Lausanne (EPFL), CH-1015 Lausanne, Switzerland}
\author{Navid Qureshi}
\affiliation{Institut Laue-Langevin, 71 Avenue des Martyrs, 38042 Grenoble Cedex 9, France}
\author{Anne Stunault}
\affiliation{Institut Laue-Langevin, 71 Avenue des Martyrs, 38042 Grenoble Cedex 9, France}
\author{Bachir Ouladdiaf}
\affiliation{Institut Laue-Langevin, 71 Avenue des Martyrs, 38042 Grenoble Cedex 9, France}
\author{J. Alberto Rodr\'iguez-Velamaz\'an}
\affiliation{Institut Laue-Langevin, 71 Avenue des Martyrs, 38042 Grenoble Cedex 9, France}
\author{YouGuo Shi}
\affiliation{Beijing National Laboratory for Condensed Matter Physics, Institute of Physics, Chinese Academy of Sciences, Beijing 100190, China}
\affiliation{School of Physical Sciences, University of Chinese Academy of Sciences, Beijing 100190, China}
\affiliation{Songshan Lake Materials Laboratory, Dongguan 523808, China}
\affiliation{Physical Science Laboratory, Huairou National Comprehensive Science Center, Beijing 101400, China}
\author{Henrik M. R{\o}nnow}
\affiliation{Institute of Physics, Ecole Polytechnique Fédérale de Lausanne (EPFL), CH-1015 Lausanne, Switzerland}
\author{Andrew T. Boothroyd}
\affiliation{Department of Physics, University of Oxford, Clarendon Laboratory, Oxford OX1 3PU, United Kingdom}
\date{\today}
\begin{abstract}
\css\, has recently been predicted to be a Weyl semimetal in which magnetic order is key to its behavior as a topological material. Here we report  unpolarized neutron diffraction and spherical neutron polarimetry measurements, supported by magnetization and transport data, which probe the magnetic order in \css\ below $T_\textrm{C} = 177$\,K. The results are fully consistent with ferromagnetic order in which the spins on the Co atoms point along the crystal $c$ axis, although we cannot rule out some canting of the spins. We find no evidence for a type of long-ranged $(\textbf{k}=\textbf{0})$ in-plane 120$^\circ$ antiferromagnetic  order which had previously been considered as a secondary phase present at temperatures between $\sim$90\,K and $T_\textrm{C}$.  A discontinuous change in bulk properties and neutron polarization observed at $T = 125$\,K when samples are cooled in a field and measured on warming is found to be due to a sudden reduction in ferromagnetic domain size. Our results lend support to the theoretical predictions that \css\, is a magnetic Weyl semimetal.
\end{abstract}
	\maketitle
	\section{Introduction}
	A key strategy in the rapidly expanding field of magnetic topological materials is the search for materials which host topological electrons that are strongly coupled to magnetic order. The general aim is to identify systems in which the topological character of the electronic bands can be controlled by altering the spin structure~\cite{suzuki_large_2016,PhysRevLett.124.076403,PhysRevB.100.201102,Wang2019,Ma2019,jin2021multiple,Ghimire2019}.
	
\begin{figure}[b!]
		\includegraphics[width=0.49\textwidth]{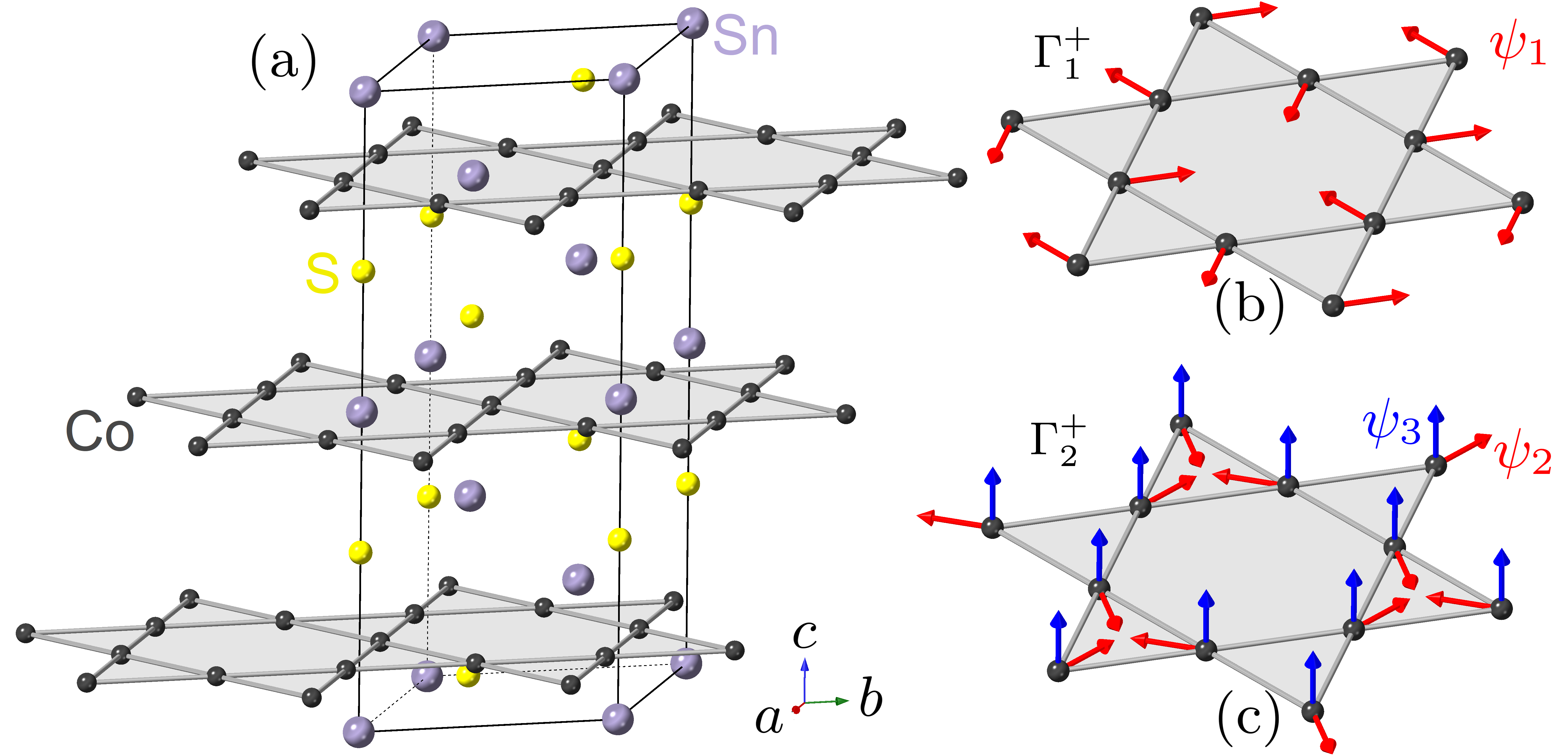}
		\caption{\label{fig:AA26_Co3Sn2S2_Figure_1.png} (a) \css\, crystallizes in the $R\bar{3}m$ space group, with the Co atoms arranged on kagome planes stacked along the crystal $c$ axis at $z = \frac{1}{6}$, $\frac{3}{6}$ and $\frac{5}{6}$. In this work we adopt the hexagonal setting to describe the unit cell, as shown. (b)--(c) Depiction of three symmetry-allowed spin configurations ($\psi_i$) of the Co moments compatible with the $\textbf{k} = \textbf{0}$ magnetic propagation vector. (b) The $\Gamma_1^+$ irreducible representation (irrep) has one basis vector, $\psi_1$, which corresponds to an in-plane $120^\circ$ AFM order. (c) The $\Gamma_2^+$ irrep, has two basis vectors which describe a different in-plane $120^\circ$ AFM order ($\psi_2$) and a FM order along the crystal $c$ axis ($\psi_3$).}
	\end{figure}

	Very recently, \css\, was proposed as a material which hosts different types of topological fermions depending on the nature of the magnetic order~\cite{Ghimire2019,yang_field-modulated_2020,RunYang2020,Zhang_2021,Liu_GiantAHE_2018}. The unit cell of \css\, can be described by the $R\bar{3}m$ space group, with the Co atoms, which reside on the $9d$ Wyckoff position, arranged on two-dimensional kagome lattices stacked along the crystal $c$ axis [see Fig.~\ref{fig:AA26_Co3Sn2S2_Figure_1.png}(a)]. Despite the wealth of reports on the topology of the electronic bands~\cite{Liu_GiantAHE_2018,LiuARPES,guguchia_tunable_2020,lachman_exchange_2020,Xu2020,liu_spin_2020,liu2021direct,liu_anisotropic_2021,Okamura_giant_2020,RunYang2020,Xu_DFT_2018,Zhang_2021,lee_moke_2021,wang_large_2018}, two fundamental questions regarding the magnetic order of the Co sublattice remain unresolved.
	
	The first concerns the magnetic structure of samples cooled below $T_\mathrm{C} = 177$\,K in zero magnetic field (ZFC--ZFW). A series of \textit{ab initio} studies~\cite{LiuARPES,liu2021direct,Xu2020,Liu_GiantAHE_2018,wang_large_2018,Okamura_giant_2020,Xu_DFT_2018,PhysRevB.88.144404} predicted half-metallic ferromagnetic (FM) order with Co moments aligned along the \textit{c} axis, Fig.~\ref{fig:AA26_Co3Sn2S2_Figure_1.png}(c). Such a spin configuration would result in an exchange splitting of the electronic bands, creating three pairs of Weyl nodes in the Brillouin zone where conduction and valence bands meet at discrete and topologically protected points in momentum space.
	
	However, a recent muon-spin rotation ($\mu$SR) study suggested that an in-plane 120$^\circ$ antiferromagnetic (AFM) order, shown in Fig.~\ref{fig:AA26_Co3Sn2S2_Figure_1.png}(b), coexists with the out-of-plane FM order in the temperature range $\sim$90\,K\,$< T <$\,177\,K~\cite{guguchia_tunable_2020}, with the AFM phase attaining a very substantial volume fraction of $80\%$ at 170\,K. There are in fact two symmetry-allowed 120$^\circ$ AFM structures, and both are predicted to generate different numbers of Weyl nodes, and at different locations in the Brillouin zone, compared with one another and with the FM state~\cite{Zhang_2021}. Moreover, if the Co moments are FM aligned perpendicular to the $c$ axis, e.g.~by an external field, then conduction and valence bands meet along a loop (nodal line) in reciprocal space~\cite{Ozawa2019}. We see, therefore, that magnetic order is closely intertwined with the topology of the electronic bands in \css, and so to understand its physical properties it is vital to know the true nature of the magnetic order.
	
	The second question concerns the anomaly observed in the vicinity of $T_\textrm{A} \simeq 125$\,K  in  magnetic, transport and optical measurements~\cite{lachman_exchange_2020,lee_moke_2021,Okamura_giant_2020}. In measurements of the Hall resistivity, Lachman \textit{et al.}~found that the anomaly becomes extremely sharp and discontinuous when the sample is cooled in a field and measurements  performed on warming in zero field (FC--ZFW)~\cite{lachman_exchange_2020}. Very recently, magneto-optical Kerr effect (MOKE) studies of \css\, discovered that this sharp transition is concomitant with
	a transformation in which a single magnetic domain  in a field-cooled sample will spontaneously break up into many smaller domains~\cite{lee_moke_2021,Okamura_giant_2020}. As magnetism is central to the topological behavior of \css, it is important to clarify how the magnetic phase evolves across the discontinuous transition at $T_\textrm{A}$ in a bulk sample.
	
	In this  work we employed unpolarised and polarised neutron diffraction along with magnetization and Hall resistivity measurements to (i) investigate the magnetic order of the Co sublattice  below $T_\textrm{C}$, and (ii) qualitatively study the evolution of the spin configuration of FC--ZFW samples in the vicinity of $T_\textrm{A}$. We find that ferromagnetic long-range order with Co moments along the crystal $c$ axis [$\psi_3$ structure in  Fig.~\ref{fig:AA26_Co3Sn2S2_Figure_1.png}(c)] can fully describe the neutron diffraction data measured in zero field, although some canting of the spins due to a coupling between the FM order and the $\psi_2$ in-plane AFM structure with the same $\Gamma_2^+$ symmetry is also possible.  On the other hand, we find no evidence for the long-ranged ($\textbf{k}=\textbf{0}$) in-plane AFM order with $\Gamma_1^+$ symmetry [$\psi_1$ structure, Fig.~\ref{fig:AA26_Co3Sn2S2_Figure_1.png}(b)] that was previously suggested to appear above $\sim$90\,K~\cite{guguchia_tunable_2020}. 
	Measurements performed by the FC--ZFW method confirm that the anomaly at $T_\textrm{A}$ is caused by a sudden reduction in the size of the magnetic domains.
	\section{Methods}
	
	Our single crystals of \css\, were grown by the self-flux method~\cite{liu_anisotropic_2021,Xu2020}, giving rise to shiny hexagonal platelets with typical dimensions $10\times10\times1$ mm$^3$. Details of the structural characterization of our crystals can be found in earlier reports~\cite{liu_anisotropic_2021,Xu2020} and in the Supplemental Material~\cite{Co3Sn2S2Supp}. Magnetization and Hall resistivity measurements were performed on a Quantum Design Physical Property Measurement System (PPMS). The temperature-dependent curves were measured from $T$ = 2 to 200\,K with a 0.5\,T field applied along the crystal $c$ axis.
	
	\begin{figure}[b!]
		\includegraphics[width=0.49\textwidth]{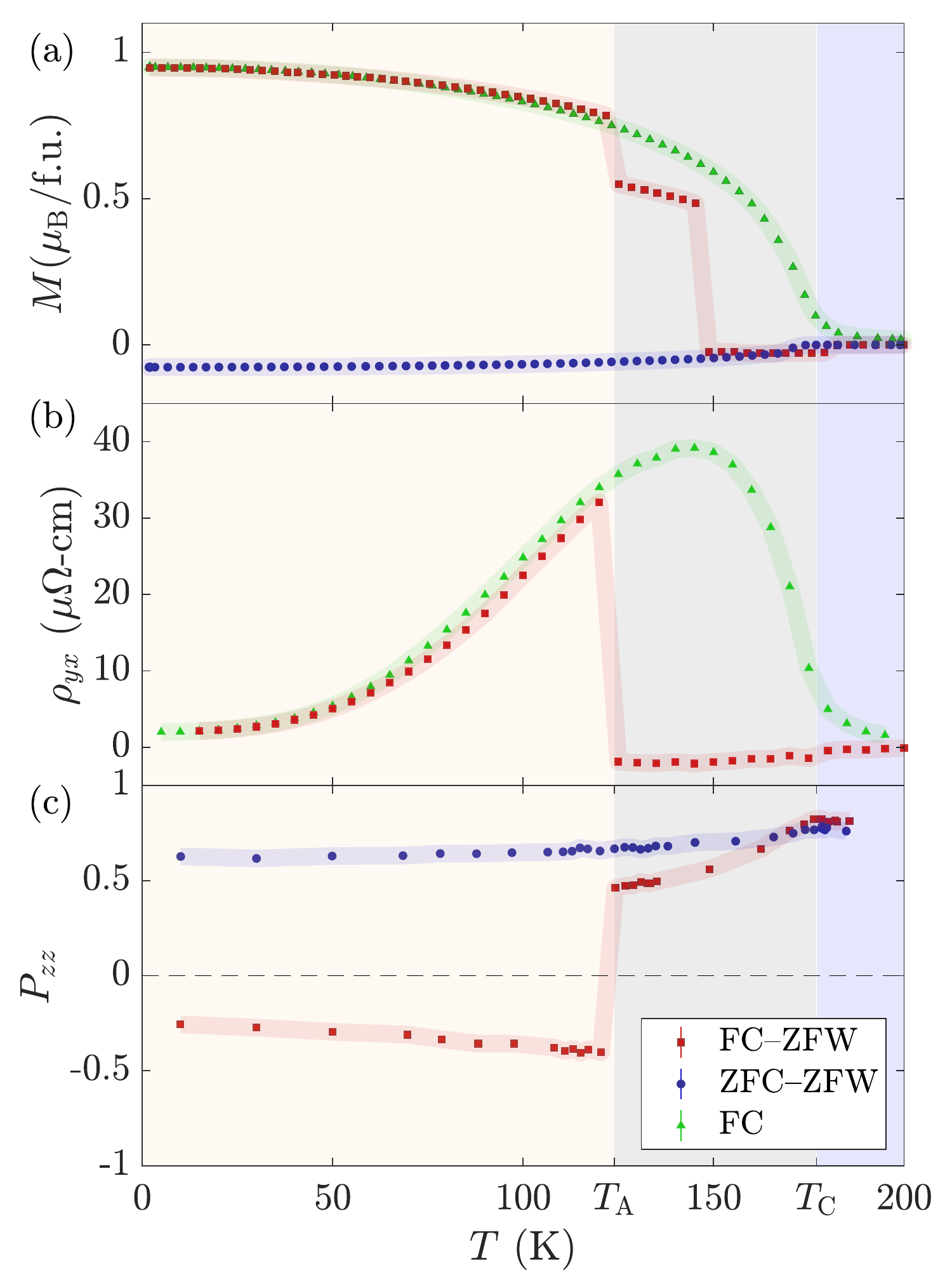}
		\caption{\label{fig:AA26_Co3Sn2S2_Figure_5} (a) Magnetization ($M$), (b) Hall resistivity ($\rho_{yx}$), and (c) $P_{zz}$ neutron polarization ($\overline{1}02$ reflection) of \css\, as a function of temperature. Three different measurement protocols were used: (red squares) the sample cooled in a field of 0.5 T along the crystal $c$ axis, and measured on warming in zero field (FC--ZFW); (blue circles) the sample cooled in zero field and measured on warming in zero field (ZFC--ZFW); (green triangles) measurements made while cooling in a field (FC).}
	\end{figure}
	
	The magnetic structure of a ZFC--ZFW \css\, sample was studied by single-crystal neutron diffraction on the four-circle diffractometer D10 at the Institut Laue--Langevin (ILL) with unpolarized neutrons ($\lambda = 2.36$\,\AA). A total of 138 reflections were measured at $T = 15$, 150 and 200\,K. In neutron powder diffraction measurements, magnetic Bragg peaks were observed to  coincide with the structural Bragg peaks, implying a \textbf{k} = \textbf{0} magnetic propagation vector~\cite{guguchia_tunable_2020}. Given, in addition, that the ordered moments are small, $\sim$0.3$\,\mu_\mathrm{B}$/Co, it is difficult to separate the weak magnetic contribution to the strong nuclear Bragg reflections with unpolarized neutrons. In a separate experiment, therefore, we employed spherical neutron polarimetry (SNP) to isolate the magnetic contribution. The SNP measurements were performed on the D3 instrument at the ILL with the CryoPAD neutron polarimeter~\cite{doi:10.1063/1.340709,TASSET1989627}. Polarised neutrons of wavelength $\lambda = 0.832$\,\AA\, were selected with a Heusler (Cu$_2$MnAl) monochromator, and the scattered beam polarization was measured with a $^3$He spin filter. Measurements of the polarization matrix $P_{ij}$ were made at several Bragg reflections for temperatures between 2 and 200\,K.   Here, the indices $i$ and $j$ indicate the directions of polarization of the incident and scattered beams, respectively, and are referred to a set of Cartesian axes defined with $x$ along the scattering vector, $z$ normal to the horizontal scattering plane, and $y$ chosen to complete a right-handed set (see Supplemental Material~\cite{Co3Sn2S2Supp}). The presented $P_{ij}$ values have been corrected for the beam polarization and spin filter efficiency as determined from a nuclear Bragg reflection with negligible magnetic component.  The sample was situated within two concentric Meissner shields to reject external magnetic fields ($B<0.1\mu$T)~\cite{doi:10.1063/1.340709,TASSET1989627}. The single crystal measured on D3 was the same as that studied on D10, and was pre-aligned on the neutron Laue diffractometer OrientExpress (ILL). Data from D3 and D10 were modelled with the \textsc{mag2pol} software~\cite{QureshiMag2Pol}.

	
	\section{Results}
	
	Figure~\ref{fig:AA26_Co3Sn2S2_Figure_5}(a) and (b) display the magnetization and Hall resistivity of \css\, as a function of temperature. The field-cooled (FC) $M(T)$ and $\rho_{yx}(T)$ curves both display an onset at $T_\mathrm{C}=177$\,K, below which $M(T)$ increases continuously to a maximum of 0.32$\mu_\mathrm{B}$/Co at $T = 2$\,K, and $\rho_{yx}$ attains a peak anomalous Hall effect near 150\,K. On warming in zero field (FC--ZFW), $M(T)$ and $\rho_{yx}(T)$ both display a sharp anomaly at $T_\textrm{A} = 125$\,K, and $M(T)$ has a second sharp drop at 150\,K. No sharp anomalies are observed for the ZFC--ZFW data. These results are in good agreement with earlier studies in which similar measurement protocols were used~\cite{lachman_exchange_2020,lee_moke_2021,Liu_GiantAHE_2018,wang_large_2018,shin2021degenerate}.  Similar behavior is observed in the neutron polarization, Fig.~\ref{fig:AA26_Co3Sn2S2_Figure_5}(c), which will be discussed later.  
	
	
\begin{figure}[t!]
\includegraphics[width=0.5\textwidth]{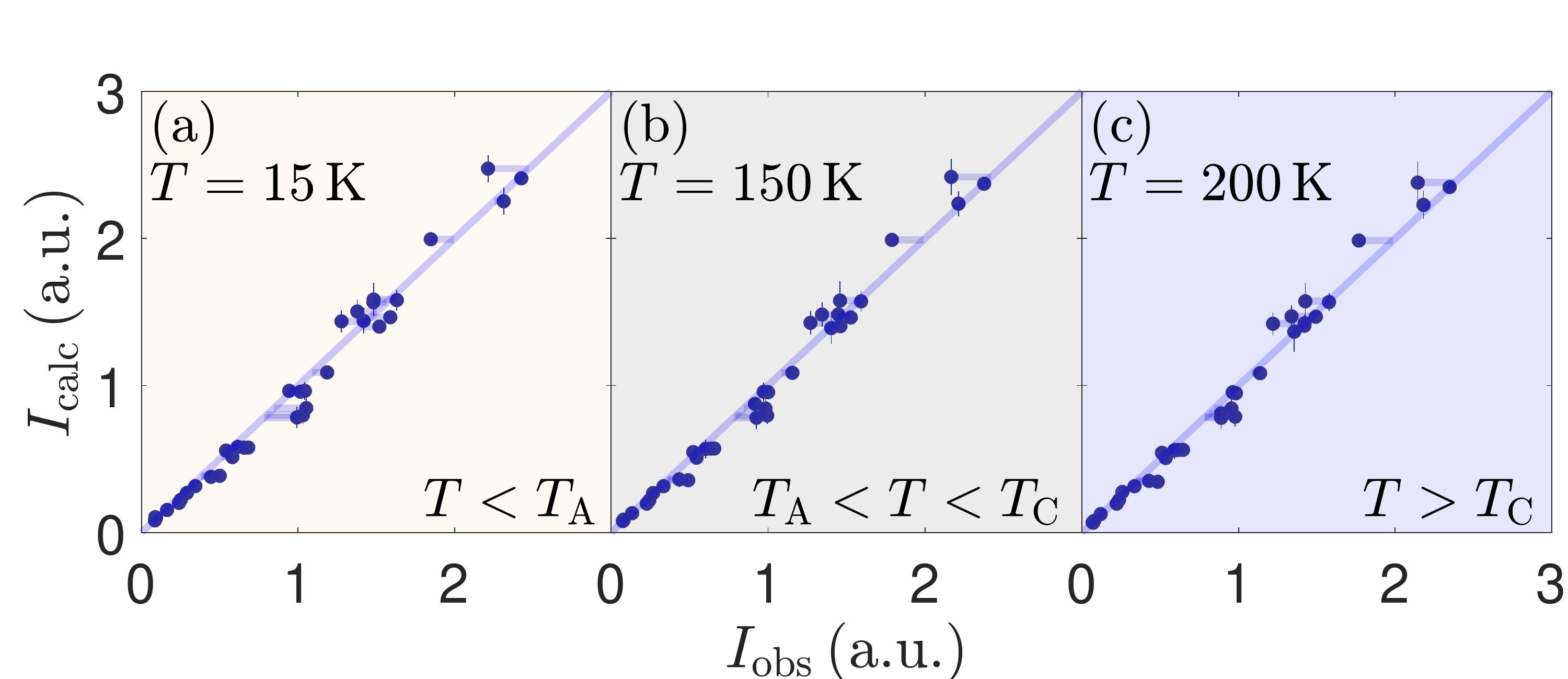}
\caption{\label{fig:AA26_Co3Sn2S2_Figure_2.png}(a)--(c) Comparison of measured ($I_\mathrm{obs}$) and calculated ($I_\mathrm{calc}$) integrated intensities from the structural and magnetic refinement in the magnetically ordered phases ($T$\,=\,15\,K and 150\,K) and the paramagnetic phase ($T$\,=\,200\,K) of \css\, respectively.}
\end{figure}
	
	Figure~\ref{fig:AA26_Co3Sn2S2_Figure_2.png} shows integrated Bragg peak intensities measured at 15, 150 and 200\,K, together with the intensities calculated from model refinements. Because the structural and magnetic diffraction peaks coincide when $T < T_{\rm C}$, the data at 200\,K in the paramagnetic phase ($T > T_{\rm C}$) provides an important point of reference. Figure~\ref{fig:AA26_Co3Sn2S2_Figure_2.png}(c) compares the measurements with the  results of a structural refinement. To account for the neutron absorption, especially that of Co ($\sigma_\mathrm{abs} \simeq 37$\,b), we corrected the integrated intensities for the attenuation along the neutron path through the crystal for each reflection using the \textsc{mag2pol} software~\cite{QureshiMag2Pol}. We find that the structural refinement provides a reasonably good fit to the data ($\chi^2_\mathrm{r}$ = 5.26; $R_\mathrm{F}$ = 3.71).
	
	We now consider the $T = 15$ and 150\,K data sets corresponding to the magnetically ordered phases of \css. The Co atoms occupy the $9d$ Wyckoff position, and the symmetry-allowed magnetic structures compatible with a $\textbf{k} = \textbf{0}$ magnetic propagation vector can be decomposed into three irreducible representations (irreps),  $\Gamma$ = $\Gamma_1^+$ + $2\Gamma_2^+$ + $3\Gamma_3^+$. These  have one, two ($2\times 1$) and six ($3\times 2$) basis vectors ($\psi_i$), respectively (see Supplemental Material~\cite{Co3Sn2S2Supp}). We refined each of the nine symmetry-allowed magnetic structures against the 15\,K and 150\,K data sets. For each structure we also refined the size of the Co moment and the magnetic domain populations, where applicable. We find that the $\psi_3$ model, with Co moments arranged ferromagnetically  along the $c$ axis [see Fig.~\ref{fig:AA26_Co3Sn2S2_Figure_1.png}(c)], provides the best fit to both data sets. Figures~\ref{fig:AA26_Co3Sn2S2_Figure_2.png}(a) and (b) show good agreement between the calculated and the measured integrated intensities at 15\,K and 150\,K, respectively. The size of the Co moment obtained from the fits is $0.39(4)\,\mu_\mathrm{B}$ at $15$\,K and $0.22(4)\,\mu_\mathrm{B}$ at 150\,K.

 Although favouring the FM structure, the unpolarized neutron study does not have sufficient sensitivity due to the relatively small moment on the Co ions to exclude a magnetic structure which combines $\psi_3$ FM order along the $c$ axis with a minority component of the in-plane $\psi_1$ AFM structure shown in Fig.~\ref{fig:AA26_Co3Sn2S2_Figure_1.png}(b). Hence, we turn to the SNP technique, which provides additional capability for a more complete characterization of the magnetic structure.

\begin{figure}[b!]
		\includegraphics[width=0.49\textwidth]{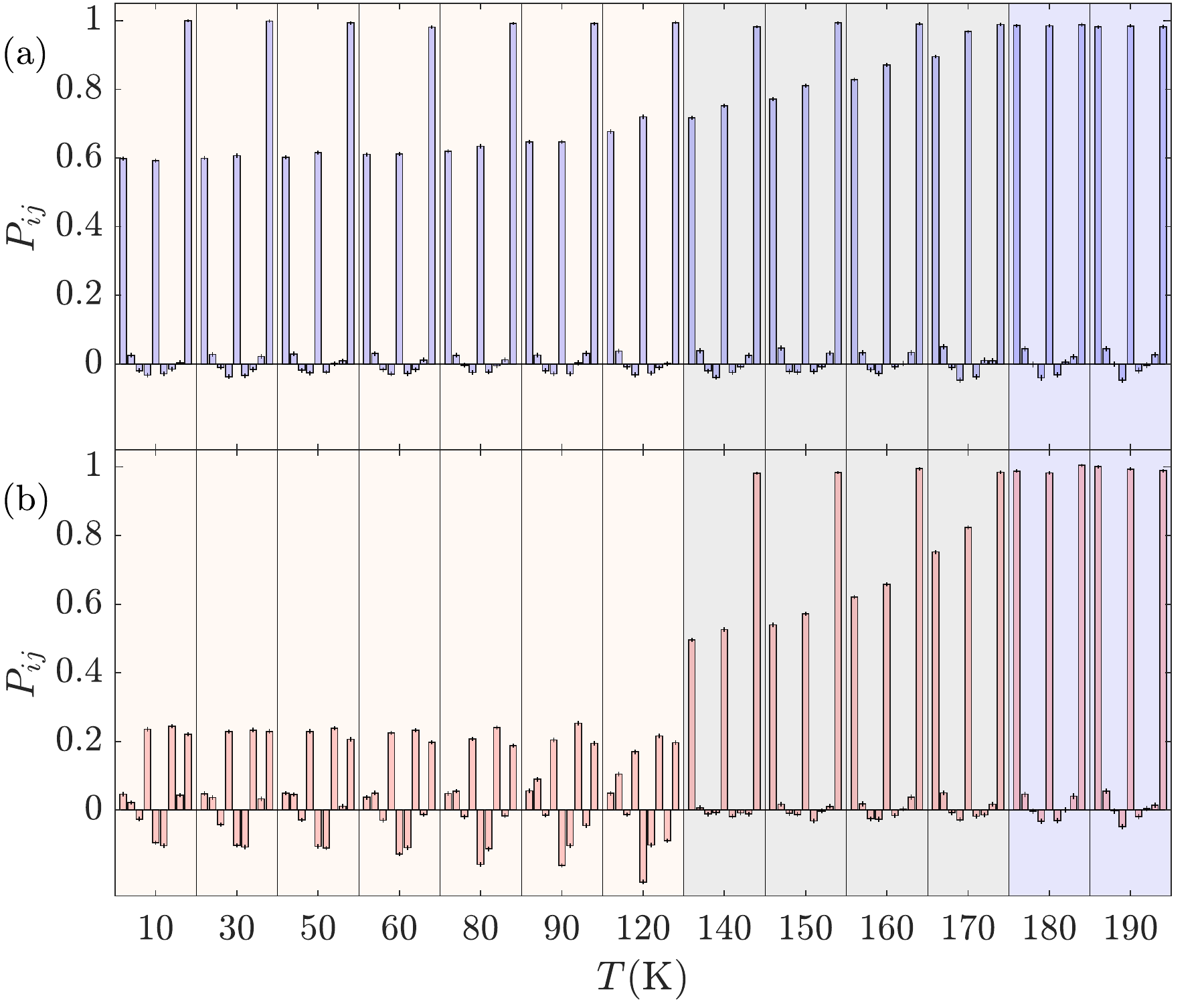}
		\caption{\label{fig:AA24h_cvert_FC_ZFC} Polarization matrices $P_{ij}$ measured at the $110$ reflection of \css\,  as a function of temperature.  (a) and (b) correspond to measurements performed with the ZFC--ZFW and FC--ZFW protocols, respectively. For each temperature, the nine matrix elements are arranged in the order (from left to right): $P_{xx}$, $P_{xy}$, $P_{xz}$, $P_{yx}$, $P_{yy}$, $P_{yz}$, $P_{zx}$, $P_{zy}$ and $P_{zz}$.}
	\end{figure}

\begin{figure*}[t!]
\includegraphics[width=0.79\textwidth]{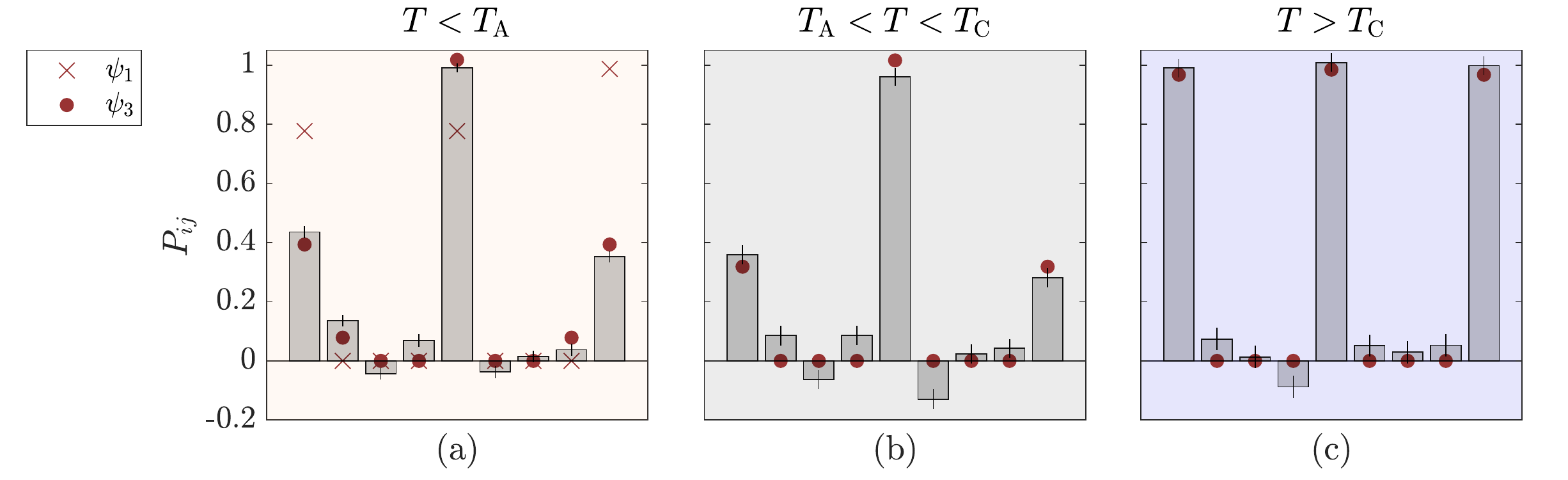}
\caption{\label{fig:AA24g_Co3Sn2S2_ZFC} Polarization matrix components $P_{ij}$ for the $101$ reflection with the $b$ axis vertical. The $P_{ij}$ are arranged in the same order on the horizontal axes as in Fig.~\ref{fig:AA24h_cvert_FC_ZFC}. The data are the vertical gray bars, and the filled red circles are calculations for the $\psi_3$ FM   structure with a ordered moment of 0.39\,$\mu_\textrm{B}$/Co. (a) $T=10$\,K. The $\times$ symbols are calculated for the $\psi_1$ AFM structure with an ordered moment of 0.1\,$\mu_\textrm{B}$/Co. (b) $T=160$\,K. (c) $T=200$\,K.}
\end{figure*}

We first consider the ZFC--ZFW measurements shown in Fig.~\ref{fig:AA24h_cvert_FC_ZFC}(a), which were measured at the 110 reflection. In the paramagnetic phase ($T>T_\textrm{C}$), the off-diagonal components of the polarization matrix ($P_{ij}$, $i\neq\,j$) are all zero and the diagonal elements ($i=j$) have $P_{ij}=1$, as expected for purely nuclear Bragg scattering (see Supplemental Material~\cite{Co3Sn2S2Supp}). For $T<T_\mathrm{C}$, $P_{zz}$ is seen to remain at $P_{zz} = 1$, so there is no measurable depolarization of the neutron beam when the neutron spins before and after scattering are parallel to $z$, which corresponds to the crystal $c$ axis for the $110$ reflection. This implies that the FM component of the sample magnetization lies along $c$, and that the volume fraction of any in-plane FM ordered moments is too small to cause a measurable depolarization. On the other hand, the $P_{xx}$ and $P_{yy}$ elements drop significantly below $T_\textrm{C}$, reaching $P_{xx} = P_{yy} \simeq 0.5$ at 10\,K. The $110$ reflection has a relatively large nuclear structure factor, such that the magnetic contribution to the diagonal $P_{ii}$ terms is negligible for the expected magnetic moment of $\sim$0.3$\,\mu_\textrm{B}$/Co. Therefore, the observed temperature dependence of $P_{xx}$ and $P_{yy}$ is almost entirely due to depolarization of the neutron beam which occurs because for $P_{xx}$ and $P_{yy}$ the neutron spins are perpendicular to the sample magnetization. The $P_{xx}$ and $P_{yy}$ measured at this reflection are therefore a proxy for the sample magnetization.

We now consider the temperature dependence of the $P_{ij}$ recorded with the FC--ZFW protocol, Fig.~\ref{fig:AA24h_cvert_FC_ZFC}(b). The initial FC state was prepared by applying an external magnetic field of 0.5\,T along the crystal $c$ axis, then cooling the sample through $T_\mathrm{C}$ down to 10\,K before inserting it into the zero-field chamber of CryoPAD. Figure~\ref{fig:AA24h_cvert_FC_ZFC}(b) shows $P_{ij}(T)$ for the $110$ reflection measured on warming in zero field. There are three distinct regimes of behavior:
	
(I) In the low temperature regime ($T<T_\textrm{A}$), we observe small values of all the $P_{ij}$. This behavior, especially the low $P_{zz}$, implies strong depolarization of the neutron beam due to an external dipolar field emanating from uncompensated ferromagnetism in the sample.
	
(II) On warming above $T= 125$\,K there is an abrupt change in the neutron polarization where the sample enters the intermediate phase [$T_\textrm{A}<T<T_\textrm{C}$, see also Fig.~\ref{fig:AA26_Co3Sn2S2_Figure_5}(c)]. In this temperature range, $P_{zz}$ recovers fully to $P_{zz} = 1$ as in the ZFC--ZFW case, Fig.~\ref{fig:AA24h_cvert_FC_ZFC}(a), indicating that the stray field outside the sample is negligible due to an equal population of FM domains with spins parallel and antiparallel to the $c$ axis. The $P_{xx}$ and $P_{yy}$ points, however, lie below the ZFC--ZFW curve (but follow a similar trend with temperature). This indicates that when the intermediate phase is reached via the FC-ZFW procedure the FM domains are larger than after the ZFC--ZFW history, giving more depolarization within the sample.

(III) In the paramagnetic phase ($T>T_\textrm{C}$) the results are the same as in the ZFC--ZFW experiment.

The SNP data recorded at the $110$ reflection shows that Co$_3$Sn$_2$S$_2$ has $c$-axis ferromagnetism, consistent with the bulk magnetization. This conclusion is reached purely from the observed depolarization of the neutron beam, and doesn't rule out the presence of other magnetic structures. To probe the magnetic structure in detail by SNP we need a Bragg peak which minimises neutron depolarisation and is sensitive to the magnetic structure factor. The $101$ reflection with the $b$ axis vertical fulfilled these requirements because the neutron path through the sample was nine times smaller than at $110$ (see Supplemental Material~\cite{Co3Sn2S2Supp}), and the size of the nuclear and magnetic structure factors at $101$ are comparable in magnitude. 

In Fig.~\ref{fig:AA24g_Co3Sn2S2_ZFC} we plot the $P_{ij}$ measured at the $101$ reflection at $T = 10$, 160 and 200\,K. In the paramagnetic phase, Fig.~\ref{fig:AA24g_Co3Sn2S2_ZFC}(c), the data are consistent with pure nuclear Bragg scattering ($P_{i=j} = 1, P_{i\ne j} = 0$), as before. On cooling below $T_\mathrm{C}$, the value of $P_{yy}$ remains almost unchanged while $P_{xx}$ and $P_{zz}$ both decrease.
The polarization matrix measured at $T = 10$\,K after ZFC, Fig.~\ref{fig:AA24g_Co3Sn2S2_ZFC}(a), is very well described by the $\psi_3$ FM  structure, which is predicted to have $P_{xx}=P_{zz}<P_{yy}=1$. On the other hand, the data are not consistent with the $\psi_1$ $120^\circ$ AFM structure which predicts $P_{xx}=P_{yy}<P_{zz}=1$ [see Fig.~\ref{fig:AA24g_Co3Sn2S2_ZFC}(a)]. The ordered moment that best fits the data in the $\psi_3$ FM structure is 0.39(2)\,$\mu_\mathrm{B}$/Co. This moment value is consistent with the result from our unpolarized neutron diffraction study but is slightly larger than the 0.3 to 0.35\,$\mu_\mathrm{B}$ usually obtained from bulk magnetization data ~\cite{PhysRevB.88.144404,Liu_GiantAHE_2018,wang_large_2018}. This may be the result of a small amount of depolarization  which slightly reduces the $P_{xx}$ and $P_{zz}$ components at the $101$ reflection and requires a larger moment to compensate. 
 
 
We note that the calculated $P_{ij}$ values at the $101$ reflection for the $\psi_2$ in-plane AFM structure are exactly the same as those for the $\psi_3$ FM structure (see Supplemental Material~\cite{Co3Sn2S2Supp}), albeit that a larger ordered moment is required for the AFM structure to give the same $P_{xx}$ and $P_{zz}$ values as the FM structure. Therefore, we cannot distinguish the $\psi_2$ AFM and $\psi_3$ FM structures with our SNP data. Further, as the $\psi_2$ and $\psi_3$ basis vectors belong to the same irrep they can, in principle, couple, leading to a canted magnetic structure. If we assume that canting does occur, and fix the $\psi_3$ FM moment to be 0.32\,$\mu_\textrm{B}$ from our magnetization data, then a fit to the 10\,K SNP data in  Fig.~\ref{fig:AA24g_Co3Sn2S2_ZFC}(a) gives a value of 0.11(4)$\mu_\textrm{B}$ for the $\psi_2$ component of the ordered moment.  However, although we cannot rule it out from our data, such a canted structure would seem unlikely given the overwhelming evidence from bulk magnetization, $\mu$SR, inelastic neutron scattering and \textit{ab initio} calculations in favour of a pure $\psi_3$ $c$-axis FM structure at low temperatures~\cite{Zhang_2021,Shen_Magnetization_2019,guguchia_tunable_2020,liu_spin_2020,PhysRevB.105.014415,Xu_DFT_2018}.

\section{Discussion}
	
The $\mu$SR study of Ref.~\onlinecite{guguchia_tunable_2020} found $\sim$100\% volume fraction of magnetically ordered phase up to near $T_\textrm{C}$, and a single muon precession frequency below $T\simeq 90$\,K consistent with a pure FM phase. Therefore, our neutron diffraction and SNP data are fully consistent with the $\mu$SR results below 90\,K.  Above $\sim$90\,K, a second frequency component was observed in the $\mu$SR time spectrum which grew with increasing temperature towards $T_\textrm{C}$. In Ref.~\onlinecite{guguchia_tunable_2020} this feature was interpreted as a second long-range ordered magnetic phase which coexists with FM order and which has the  $\psi_1$ AFM structure shown in Fig.~\ref{fig:AA26_Co3Sn2S2_Figure_1.png}(b). The volume fraction of the proposed in-plane AFM order grows from $0\%$ at $T=90$\,K to $80\%$ at $T=170$\,K, which at such high volume fraction can be considered as long-range (\textbf{k}=\textbf{0}) order. This interpretation, however, is not consistent with our SNP data.

First, the $\psi_1$ AFM structure does not describe the SNP data in Fig.~\ref{fig:AA24g_Co3Sn2S2_ZFC}(b), which shows the $P_{ij}$  in the intermediate phase at $T = 160$\,K measured with the FC--ZFW protocol~\cite{footnote1}. The  $P_{ij}$ at $160$\,K are similar to those at 10\,K, and the FM structure gives a good description of both sets of data.  Note that for a pure FM phase, $P_{xx}$ and $P_{zz}$ are expected to be larger at $160$\,K than at 10\,K  due to the decrease in the ordered moment. The fact that this is not observed is likely due to neutron depolarization which, as mentioned earlier, is stronger for a FC sample than a ZFC sample. Nevertheless, depolarization is expected to affect the $P_{xx}$ and $P_{zz}$ components by a similar amount (and more so than $P_{yy}$, since in this measurement the $y$ axis is nearly parallel to the crystal $c$ axis which is along the sample magnetization), and so it is reasonable to compare (qualitatively, at least) the measured $P_{ii}$ with a model. On this basis we find no evidence in the data for a significant amount of the $\psi_1$ structure, since the latter predicts $P_{xx} = P_{yy} < P_{zz} = 1$ whereas the data have $P_{xx} \simeq P_{zz} < P_{yy}$.

\begin{figure}[t!]
\includegraphics[width=0.49\textwidth]{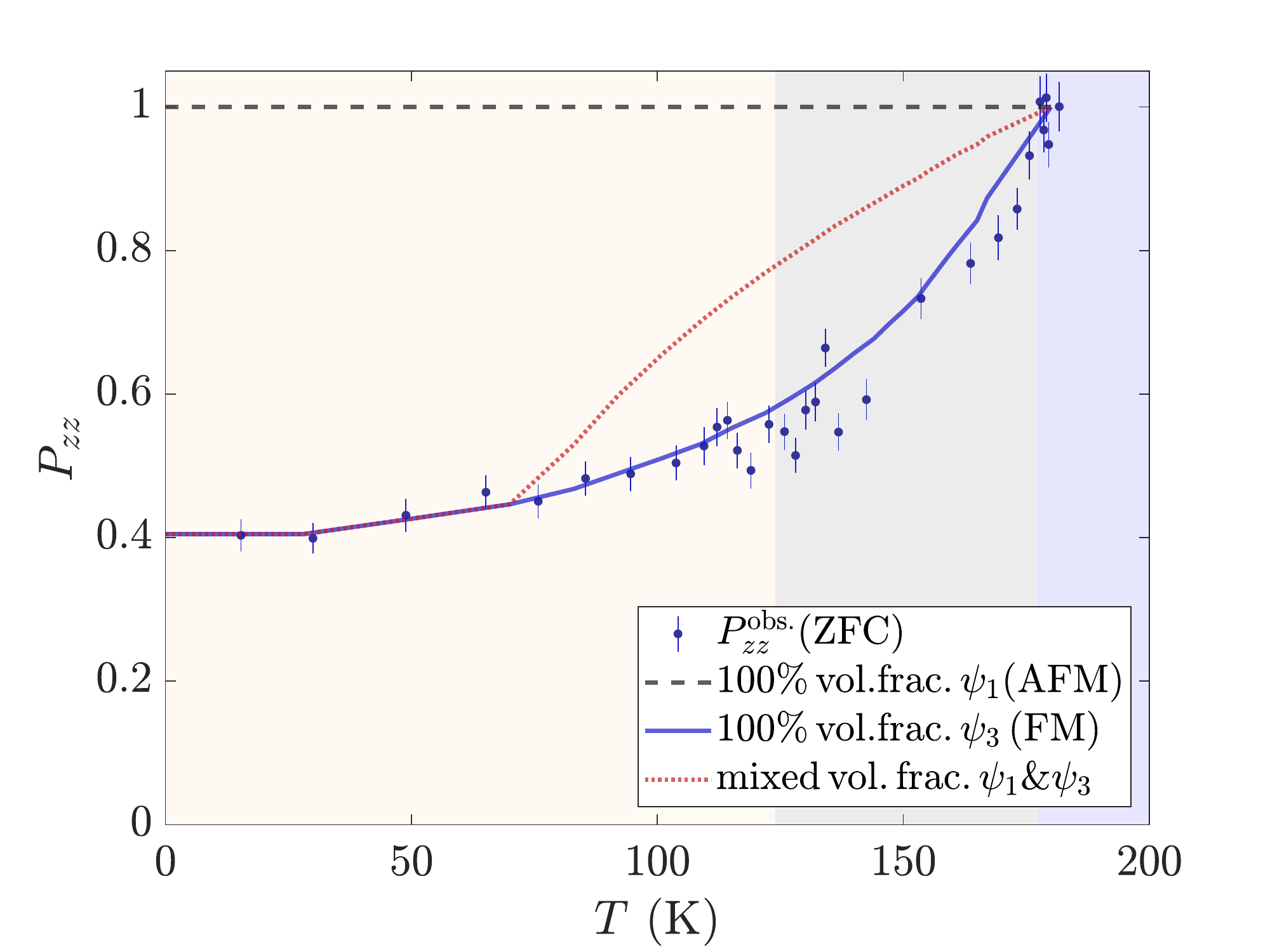}
\caption{\label{fig:AA24b_Co3Sn2S2_101} Temperature dependence of the $P_{zz}$ matrix element of \css\, measured at the 101 reflection with the ZFC--ZFW protocol. The black dashed, solid blue and red dotted lines denote the calculated $P_{zz}$ assuming, respectively, $100\%$  in-plane $\psi_1$ AFM order,  $\psi_3$ $100\%$ FM order, and a coexistence of $\psi_1$ and $\psi_3$  in the ratio inferred by $\mu$SR in Ref.~\cite{guguchia_tunable_2020}. Details of the calculation of $P_{zz}$ can be found in the Supplemental Material~\cite{Co3Sn2S2Supp}.}
\end{figure}

Second, Fig.~\ref{fig:AA24b_Co3Sn2S2_101} compares the temperature dependence of $P_{zz}$ measured at the 101 reflection by the ZFC--ZFW protocol with  $P_{zz}$ calculated for three cases: (I) pure $\psi_1$ in-plane AFM order, (II) pure $\psi_3$ FM order, and (III) a coexistence of the two in the (temperature-dependent) ratio inferred from $\mu$SR by Guguchia \textit{et al.}~\cite{guguchia_tunable_2020}. The temperature dependence of the ordered moment is assumed to be proportional to the bulk magnetization. The results can be fully described by a pure FM order with the moments along $c$, and do not support the presence of a significant amount of long-range $\psi_1$ AFM order. 

As discussed earlier, our results cannot distinguish between the $\psi_2$ in-plane AFM structure and the $\psi_3$ $c$-axis FM structure, and a coexistence of these two phases might be one way to account for the second frequency component observed in the $\mu$SR spectra at temperatures between $\sim$90\,K and $T_\textrm{C}$. However, another possible reconciliation of the $\mu$SR and SNP results at intermediate temperatures can be found in a very recent study of FM domain wall dynamics in \css\, by scanning magneto-optical Kerr microscopy (MOKE)~\cite{lee_moke_2021}. Therein, Lee \textit{et al.} reports that the volume fraction of domain walls grows substantially on warming and dominates at $T$=170\,K, concomitant with the observation of a broad peak centered at $\sim$ 80 mT in FFT of the $\mu$SR spectra which was attributed by Guguchia \textit{et al.}\ to an in-plane $\psi_1$ AFM order with an $80\%$ volume fraction. Muons implanted in the vicinity of a domain wall will see a distribution of internal magnetic fields due to the change in the size and orientation of the magnetic moments across the wall. The average field in a domain wall will be less than that in the interior of a FM domain, consistent with the second $\mu$SR component~\cite{guguchia_tunable_2020}, and the reported increase in volume fraction of the phase responsible for the second frequency component as $T$ approaches $T_\textrm{C}$ is explained by the increase in the density and width of linear domain walls observed in the MOKE maps~\cite{lee_moke_2021}. As domain walls do not scatter neutrons into sharp Bragg peaks, this interpretation might be able to explain why SNP is largely insensitive to their presence. It would be interesting to model the field distribution in a domain wall and compare it with that from $\mu$SR in order to test this interpretation.

Another recent experimental finding that needs to be explained in view of the present results is the observation of exchange bias effects at $T < T_\textrm{A}$~\cite{lachman_exchange_2020}. Exchange bias appears as a shift along the field axis of the FM hysteresis loop, usually associated with the interface between a FM and an AFM. Lachman \textit{et al.}~propose that its origin in \css\, is due to the presence of a frustration-induced spin-glass phase which coexists with FM order~\cite{lachman_exchange_2020}. However, in addition to the fact that our results, and those of other studies, are consistent with pure FM order at $T < T_\textrm{A}$, an analysis of the spin-wave spectrum of \css\, did not find evidence for strong AFM interactions that could cause frustration in the kagome layers~\cite{Zhang_2021}. An alternative explanation could be that the exchange bias effect is caused by the existence of a small concentration of secondary FM domains that require a larger field to reverse than the primary FM matrix, perhaps due to disorder. In this scenario, exchange bias would be the result of the pinning interaction across the interface between the primary and secondary domains. 

Finally, we note that the dramatic recovery of neutron polarization on warming above $\sim$120\,K after cooling in a field [Fig.~\ref{fig:AA26_Co3Sn2S2_Figure_5}(c)], which is explained by a large reduction in FM domain size, correlates with the discontinuous change in magnetic, transport and optical data previously observed at $T=T_\textrm{A}$~\cite{Shen_Magnetization_2019,Kassem_2020,Kassem_2017,Wu_Magnetization_2020,Zhang_2021,shin2021degenerate} [see also Fig.~\ref{fig:AA26_Co3Sn2S2_Figure_5}(a),(b)]. The domain size transition at $T_\textrm{A}$ has recently been imaged directly in the MOKE study~\cite{lee_moke_2021}, but our SNP measurements prove that the transition occurs in the bulk, where the Weyl fermions live, as well as at the surface.

\section{Conclusion}
	
In conclusion, the central findings of this study of magnetic order in \css\ are, (1) that FM long-range order with moments along the $c$ axis [$\psi_3$ structure in Fig.~\ref{fig:AA26_Co3Sn2S2_Figure_1.png}(c)] accounts very well for our  results  at all temperatures below $T_\textrm{C}$, although we cannot exclude the possibility that the spins cant away from the $c$ axis at intermediate temperatures to form an additional component with in-plane AFM order [$\psi_2$ in Fig.~\ref{fig:AA26_Co3Sn2S2_Figure_1.png}(c)], and (2) that a sudden reduction in FM domain size takes place in FC samples on warming through $T_\textrm{A} = 125$\,K and is responsible for discontinuities in several macroscopic physical properties.
The results clarify the interplay between magnetic order and electronic band topology in \css, and emphasize the importance of understanding the highly unusual behavior of the FM domains. 
\section{acknowledgments}	
	\begin{acknowledgments}
		The authors  wish  to  thank Sebastien Vial for valuable technical help and Bruce Normand for discussions. The proposal numbers for the \css\, neutron scattering experiments are 5-41-1113 (D3, D10) and EASY-885 (OrientExpress), with the data publicly available at~\cite{D3_data_2021}. A.T.B. was supported by the Oxford--ShanghaiTech collaboration project. J.-R. Soh acknowledges support from the Singapore National Science Scholarship, Agency for Science Technology and Research. 
	\end{acknowledgments}
	\bibliography{ref.bib}
\end{document}